\documentclass[pra,aps,twocolumn,superscriptaddress,floatfix,showpacs,longbibliography]{revtex4-2}
\usepackage{graphicx}
\usepackage{dcolumn}
\usepackage{bm}
\usepackage{upgreek}
\usepackage{natbib}
\usepackage[normalem]{ulem}
\usepackage{amsmath,amssymb}
\usepackage{dsfont}
\usepackage[english]{babel}
\usepackage{color}

\begin{document}
\title{Chaotic dynamics of Bose-Einstein condensate in a density-dependent gauge field}

\author{Lei Chen}
\affiliation{Guangdong Provincial Key Laboratory of Quantum Engineering and Quantum Materials, School of Physics and Telecommunication Engineering, South China Normal University, Guangzhou 510006, China}
\affiliation{School of Physics and Electronic
Science, Zunyi Normal University, Zunyi 563006, China}

\author{Qizhong Zhu}
\email{qzzhu@m.scnu.edu.cn}
\affiliation{Guangdong Provincial Key Laboratory of Quantum Engineering and Quantum Materials, School of Physics and Telecommunication Engineering, South China Normal University, Guangzhou 510006, China}
\affiliation{Guangdong-Hong Kong Joint Laboratory of Quantum Matter, Frontier Research Institute for Physics, South China Normal University, Guangzhou 510006, China}

\date{\today}

\begin{abstract}
In this work we study the
effect of density-dependent gauge field on the collective
dynamics of a harmonically trapped Bose-Einstein condensate,
beyond the linear response regime. The density-dependent gauge field,
as a backaction of the condensate, can in turn affect the condensate dynamics, resulting in
highly nonlinear equations of motion.
We find that the dipole and breathing oscillations of the condensate along the
direction of gauge field are coupled by this field.
For a quasi-one-dimensional condensate, this coupling makes
the collective motion quasiperiodic. While for a quasi-two-dimensional condensate,
the gauge field can also induce a Hall effect, manifested as
an additional coupling between dipole and breathing oscillations in perpendicular direction.
When the density-dependent
gauge field is strong, the interplay between these oscillations can cause the collective dynamics of the condensate to become chaotic. 
Our findings reveal an important effect of dynamical gauge field on the nonlinear dynamics of a Bose-Einstein condensate.
\end{abstract}
\maketitle

\section{introduction}

The emulation of gauge field in cold atomic gases is one
of the major topics attracting persistent interest. Significant progress has
been made in the engineering of artificial gauge field in the last
decades, including both the abelian \cite{lin_synthetic_2009,lin_Synthetic_2011,dalibard_Colloquium_2011,goldman_Lightinduced_2014} and non-abelian gauge field, in particular, the
spin-orbit coupling \cite{lin_Spinorbitcoupled_2011,wang_SpinOrbit_2012,cheuk_SpinInjection_2012,goldman_Lightinduced_2014,zhai_Degenerate_2015,huang_Experimental_2016,wu_Realization_2016}. Nevertheless, most research has previously 
mainly focused on the realization
of a static gauge field, where the generated artificial gauge field is completely
determined by the external laser field, and itself
has no dynamics. Recently, simulating
dynamical gauge field in cold atoms has gained increasing interest and attention in the community. Until now, a variety of theoretical
proposals to create dynamical gauge field have been put forward and 
some already been experimentally realized \cite{keilmann_Statistically_2011,banerjee_Atomic_2012,zohar_ColdAtom_2013,tagliacozzo_Simulation_2013,
edmonds_Simulating_2013,greschner_DensityDependent_2014,ballantine_Meissnerlike_2017,
clark_Observation_2018,
schweizer_Floquet_2019,gorg_Realization_2019,kroeze_Dynamical_2019,xu_Densitydependent_2021}. 
Those schemes either exploit the
quantum nature of external control light, or make the gauge field
dependent on the atomic density.
It is an important and interesting topic to explore the new physics brought by those 
dynamical gauge fields \cite{dong_Cavityassisted_2014,zheng_Topological_2015,raventos_Topological_2016,zheng_Superradiance_2016,yao_Manybody_2020}.

One prominent effect induced by a static gauge field, e.g., the magnetic field, is the celebrated Hall effect for electrons in solids. For cold neutral gases, a superfluid can also exhibit Hall effect, when subject to an static
artificial magnetic field \cite{zhu_Spin_2006,leblanc_Observation_2012,choi_Observation_2013}. Will the dynamical gauge field also induce a Hall effect on the transverse motion of a condensate? If true, how will the collective dynamics of the condensate influenced by the Hall effect?

In this paper, we try to address these issues by considering dynamical gauge field which depends on the atomic density \cite{edmonds_Simulating_2013}. 
Previous studies on the effect of density-dependent gauge field mainly
focus on the quasi-one-dimensional (quasi-1D) case \cite{edmonds_Simulating_2013,edmonds_Elementary_2015,dingwall_Nonintegrable_2018,dingwall_Stability_2019} and vortex related physics in quasi-two-dimensional (quasi-2D) case \cite{butera_Vortex_2016,edmonds_Vortex_2020}. To
explore physics like Hall effect, the condensate dynamics has to be treated as quasi-2D.
In particular, we study the effect of density-dependent
gauge field on the collective dynamics of a Bose-Einstein condensate (BEC) confined in a harmonic trap, beyond the linear response regime. 
We find that, in quasi-1D
case, the dipole and breathing oscillations in the direction of gauge field are coupled, resulting in
quasiperiodic motion of these modes, even when the gauge field is quite
strong. While in quasi-2D case, these two modes are additionally coupled with the breathing 
oscillation in the perpendicular direction through the Hall effect, and their
interplay can lead to a chaotic dynamics in the presence of a strong gauge field. Our findings shed light on the intriguing effect
of dynamical gauge field on the nonlinear dynamics of BEC. 

\begin{figure}
\centering
\includegraphics[width=0.95\linewidth]{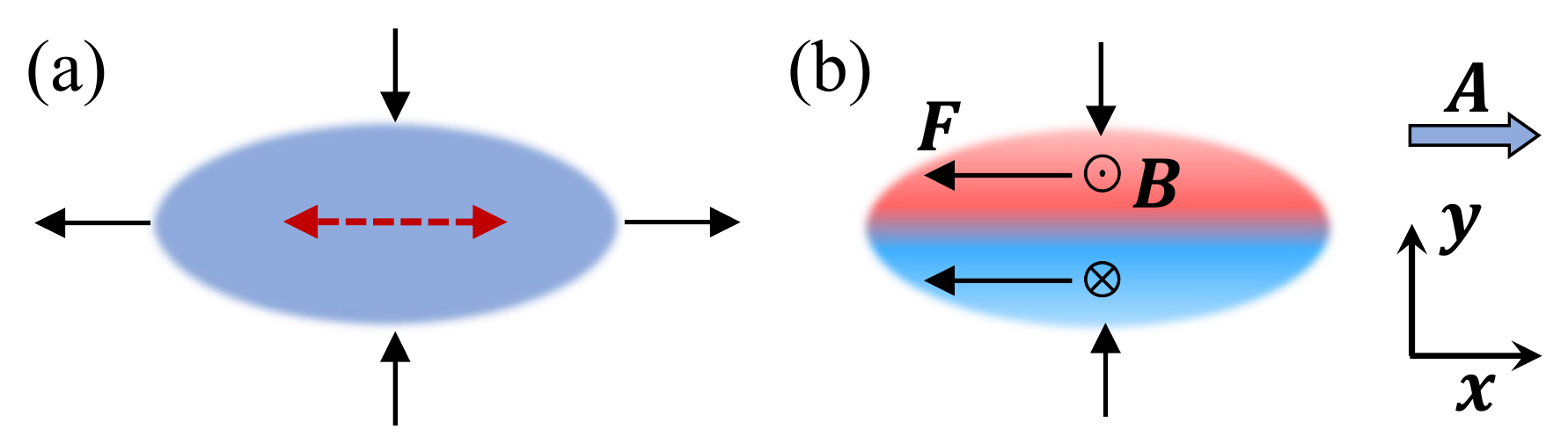}
\caption{(a) Relevant collective modes of a BEC confined in a quasi-2D harmonic trap. The density-dependent gauge field
is fixed in $x$ direction, and couples the dipole oscillation in $x$ direction (red dashed arrow) with breathing oscillations both in $x$ and $y$ directions (black arrow)(see Eq. \ref{eom}). The dipole oscillation in $y$ direction is decoupled in this scenario. (b) The Hall effect induced by the density-dependent gauge field. Red and Blue color denote the region where the effective magnetic field $\mathbf{B}$ points to $z$ and $-z$ directions, respectively. The breathing oscillation in
$y$ direction can induce a dipole oscillation in $x$ direction and vice versa, through the Lorentz force $\mathbf{F}$. \label{fig:sketch}}
\end{figure}

This paper is organized
as follows. First, in Sec. \ref{variational} we employ the variational wave function method to derive the coupled motion of center of mass and width, namely, the dipole mode and breathing mode of a BEC, and arrive at a set of nonlinear equations. 
We then solve these coupled equations numerically, with results presented in Sec. \ref{dynamics}, and investigate the nature of the condensate dynamics, namely, regular or chaotic. This is demonstrated by computing the Poincar\'e section,
both in four-dimensional (4D) form and its 2D projection \cite{lukes-gerakopoulos_Dynamics_2016}, along with the analysis of power spectrum of the dynamical variables. Finally, we end our discussion with a summary in
Sec. \ref{conclusion}.

\section{Variational wave function method}\label{variational} 

We start from the mean-field Hamiltonian which describes the atomic motion
and atom-light coupling 
\begin{equation}
\hat{H}=\left(\frac{\hat{\mathbf{p}}^{2}}{2m}+{V}(\mathbf{r})\right)\otimes\mathds{1}+{H}_\mathrm{int}+{U}_\mathrm{AL},\label{eq:1}
\end{equation}
where 
\begin{equation}
{U}_\mathrm{AL}=\frac{\hbar\Omega}{2}\left(\begin{array}{cc}
0 & e^{-i\phi\left(\mathbf{r}\right)}\\
e^{i\phi\left(\mathbf{r}\right)} & 0
\end{array}\right)\label{eq:2}
\end{equation}
describes the coupling between two internal states $\left|1\right\rangle $
and $\left|2\right\rangle $, characterized by the two-photon
Rabi frequency $\Omega$ and the laser phase $\phi\left(\mathbf{r}\right)$. 
Here ${V}(\mathbf{r})$ is the external trap potential, $\mathds{1}$ is the identity
matrix defined in the pseudospin space spanned by $\left|1\right\rangle $
and $\left|2\right\rangle $, and $H_\mathrm{int}=\left(1/2\right)\mathrm{diag}\left[g_{11}\rho_{1}+g_{12}\rho_{2},g_{22}\rho_{2}+g_{12}\rho_{1}\right]$
is the mean-field interaction characterized by two-body
interaction strengths $g_{ll^{\prime}}=4\pi\hbar^{2}a_{ll^{\prime}}/m$,
with $a_{ll^{\prime}}$ being the scattering lengths for collisions
between the components $l$ and $l^{\prime}$. 
The population of the $i$-th state is $\rho_{i}=\left|\Psi_{i}\right|^{2}$
$\left(i=1,2\right)$.

For a dilute BEC, the coupling strength
$\hbar\Omega$ is typically much larger than the mean-field energy.
Therefore we can construct the interacting dressed states through perturbation
theory with respect to the eigenstates of the atom-light coupling $U_\mathrm{AL}$,
 namely, diagonalizing $U_\mathrm{AL}+H_\mathrm{int}$
by treating $H_\mathrm{int}$ as a small perturbation.
We finally obtain the eigenstates of $U_\mathrm{AL}+H_\mathrm{int}$ represented
by perturbed dressed states $|\chi_{\pm}\rangle=|\chi_{\pm}^{(0)}\rangle+|\chi_{\pm}^{(1)}\rangle$,
where 
\begin{equation}
|\chi_{\pm}^{(1)}\rangle=\pm\frac{g_{11}-g_{22}}{8\hbar}\rho_{\pm}|\chi_{\mp}^{(0)}\rangle,\label{eq:3}
\end{equation}
with eigenvalues $g\rho_{\pm}\pm\hbar\Omega/2$. Here $g=\left(g_{11}+g_{22}+2g_{12}\right)/4$,
$\rho_{\pm}=\left|\Psi_{\pm}\right|^{2}$ and the unperturbed dressed
states $|\chi_{\pm}^{(0)}\rangle=(|1\rangle\pm\exp\{i\phi(\mathbf{r})\}|2\rangle)/\sqrt{2}$.
In order to derive an interacting gauge theory, we first expand
a general state as $\left|\xi\right\rangle =\sum_{i=\left\{ +,-\right\} }\Psi_{i}\left(\mathbf{r},t\right)\left|\chi_{i}\right\rangle $. For atomic motion slow enough, we can make the adiabatic approximation
and project the general wave function into one of the
two dressed states, $\left|\chi_{\pm}\right\rangle$. The resulting effective
Hamiltonian reads 
\begin{equation}
\hat{H}^\mathrm{eff}_{\pm}=\frac{1}{2m}\left(\hat{\mathbf{p}}-\mathbf{A}_{\pm}\right)^{2}+V\left(\mathbf{r}\right)+W\pm\frac{1}{2}\hbar\Omega+\frac{g}{2}\rho_{\pm},\label{eq:4}
\end{equation}
where a scalar potential $W=\hbar^{2}\left|\left\langle \chi_{-}|\nabla\chi_{+}\right\rangle \right|^{2}/2m$
and a geometric vector potential $\mathbf{A}_{\pm}=i\hbar\left\langle \chi_{\pm}|\nabla\chi_{\pm}\right\rangle $
are introduced. According to aforementioned
definition, the vector potential associated with the perturbed dressed state,
to leading order, is given by 
\begin{equation}
\mathbf{A}_{\pm}=\mathbf{A}^{\left(0\right)}\pm\mathbf{a}_{1}\left|\Psi_{\pm}\left(\mathbf{r}\right)\right|^{2}.\label{eq:5}
\end{equation}
Here $\mathbf{A}^{\left(0\right)}=-\left(\hbar/2\right)\nabla\phi\left(\mathbf{r}\right)$
is the single-particle vector potential and $\mathbf{a}_{1}=\left(\nabla\phi\left(\mathbf{r}\right)\right)\left(g_{11}-g_{22}\right)/\left(8\Omega\right)$
controls the effective strength of density-dependent vector potential.
In the following we only consider one branch of the dressed states,
e.g., the $+$ branch without loss of generality, and thus we can drop the
$\pm$ index in Eq. \ref{eq:4}. Note that an effective magnetic field is associated with
this density-dependent gauge field through $\mathbf{B}=\nabla\times\mathbf{A}=\nabla\rho\times\nabla\phi\left(\mathbf{r}\right)\left(g_{11}-g_{22}\right)/\left(8\Omega\right)$. So the density variation of BEC
results in a non-vanishing density-dependent magnetic field, which is absent in the
non-interacting case for the present setup. This effective magnetic field may induce a Hall effect,
and has significant influence on the collective dynamics of BEC.

It was shown that BEC subject to this 
density-dependent gauge field 
obeyed a generalized Gross-Pitaevskii equation with current nonlinearity \cite{edmonds_Simulating_2013}, and a variety of
intriguing phenomena were predicted based on this equation \cite{edmonds_Simulating_2013,edmonds_Elementary_2015,edmonds_Vortex_2020}. Here we focus on the effect
of this gauge field on the collective dynamics of BEC in a harmonic
trap. Most
studies on this topic usually consider a quasi-1D BEC, but as will be shown in the following, we find that the motion of a quasi-2D BEC
is endowed with radically new physics. To illustrate
this, we examined the dynamics of a quasi-2D BEC with motion in $z$ direction frozen out. 
By using the method of variational wave function, the equation of motion is obtained
by minimizing the Lagrangian 
\begin{equation}
\mathcal{L}=\langle\Psi|i\hbar\partial_{t}-\hat{H}^\mathrm{eff}|\Psi\rangle.\label{eq:6}
\end{equation}
In particular, we choose a well-established variational wave function \cite{perez-garcia_Low_1996}
\begin{equation}
\Psi\left(\mathbf{r},t\right)=f\left(z\right)C\left(t\right)\prod_{\eta=x,y}e^{-\frac{\left[\eta-\eta_{0}\left(t\right)\right]^{2}}{2w_{\eta}^{2}\left(t\right)}+i\eta\alpha_{\eta}\left(t\right)+i\eta^{2}\beta_{\eta}\left(t\right)}.\label{eq:7}
\end{equation}
Here the harmonic potential is given by $V(\mathbf{r})=(m\nu^{2}\lambda_{x}^{2}x^{2}+m\nu^{2}\lambda_{y}^{2}y^{2}+m\nu^{2}\lambda_{z}^{2}z^{2})/2$, with
the quasi-2D condition $\lambda_{z}\gg\lambda_{i}\left(i=x,y\right)$.
 $f(z)=\exp[-z^{2}/(2a_{z}^{2})]/(\sqrt{\pi}a_{z})^{1/2}$
represents the BEC wave function in $z$ direction, with $a_{z}=[\hbar/(m\nu\lambda_{z})]^{1/2}$ being the width of the Gaussian wave packet.
The dynamical variables are the center of mass $\eta_0=\left(x_{0},y_{0}\right)$,
 amplitude $C\left(t\right)$, condensate
width $w_{\eta}\left(t\right)$, slope $\alpha_{\eta}\left(t\right)$,
and variables related to curvature $\beta_{\eta}\left(t\right)$.
Plugging the variational wave function $\Psi\left(\mathbf{r},t\right)$ into Eq. \ref{eq:6} readily leads to the effective Lagrangian
\begin{widetext}
\begin{align}
\frac{\mathcal{L}}{N} & =\sum_{\eta=x,y}\Bigg\{\frac{\hbar}{2}\left(\dot{\beta}_{\eta}w_{\eta}^{2}+2\dot{\alpha}_{\eta}\eta_{0}+2\dot{\beta}_{\eta}\eta_{0}^{2}\right)+\frac{\hbar^{2}}{2m}\left[\frac{1}{2w_{\eta}^{2}}+\alpha_{\eta}^{2}+4\eta_{0}\alpha_{\eta}\beta_{\eta}+2\left(w_{\eta}^{2}+2\eta_{0}^{2}\right)\beta_{\eta}^{2}\right]\nonumber\\
  & -\left(\frac{\hbar}{m} A_{\eta}^{\left(0\right)}+\frac{N\hbar a_{1\eta}}{2\sqrt{2\pi^{3}}ma_{z}w_{x}w_{y}}\right)\left(\alpha_{\eta}+2\eta_{0}\beta_{\eta}\right)+\frac{1}{4}m\omega_{\eta}^{2}\left(w_{\eta}^{2}+2\eta_{0}^{2}\right)\Bigg\}+\frac{\tilde{g}N}{4\sqrt{2\pi^{3}}a_{z}w_{x}w_{y}}\nonumber\\
 & +\frac{N^{2}\mathbf{a}_{1\perp}^{2}}{6\sqrt{3}m\pi^{3}a_{z}^{2}w_{x}^{2}w_{y}^{2}},
\end{align}
\end{widetext} 
with $\tilde{g}/2=g/2+\mathbf{A}_{\perp}^{(0)}\cdot\mathbf{a}_{1\perp}/m$, $\mathbf{A}_{\perp}^{(0)}=(A_{x}^{(0)}, A_{y}^{(0)})$ and $\mathbf{a}_{1\perp}=(a_{1x}, a_{1y})$.
The dot over the variable denotes time derivative of that variable. 
In terms of the notation $q_{i}\equiv\{w_{x},w_{y},x_{0},y_{0},\alpha_{x},\alpha_{y},\beta_{x},\beta_{y}\}$,
the Lagrangian equation reads
\begin{equation}
\frac{d}{dt}\left(\frac{\partial\mathcal{L}}{\partial\dot{q}_{i}}\right)-\frac{\partial\mathcal{L}}{\partial q_{i}}=0,\label{eq:8}
\end{equation}
from which one eventually arrives at the equations of motion of those dynamical variables.

For computational simplicity and without loss of generality, we assume the laser phase takes the form of a plane wave along $x$ direction, namely, $\phi(\mathbf{r})=kx$. Thus, the density-dependent
gauge field is also along $x$ direction.
After introducing dimensionless variables and constants
according to $\tau=\nu t$, $R_{\eta}=w_{\eta}/l_{0}$ ($\eta=x,y$), $R_{z}=a_{z}/l_{0}$,
$P_x=k(g_{11}-g_{22})N/(8\Omega\hbar l_{0}^{2})$, $P_y=0$, and $G=gN/\left(\hbar\nu l_{0}^{3}\right)$,
where $l_{0}=\left[\hbar/\left(m\nu\right)\right]^{1/2}$, the resulting
equations of motion finally read as follows:
\begin{gather}
\frac{d^{2}\eta_{0}}{d\tau^{2}}+\lambda_{\eta}^{2}\eta_{0}=\frac{P_{\eta}}{2\sqrt{2\pi^{3}}R_{z}R_{x}R_{y}}\sum_{\xi=x,y}\frac{1}{R_{\xi}}\frac{dR_{\xi}}{d\tau},\nonumber\\
\frac{d^{2}R_{\eta}}{d\tau^{2}}+\lambda_{\eta}^{2}R_{\eta}=\frac{1}{R_{\eta}^{3}}
+\left(\frac{2}{3\sqrt{3}}-\frac{1}{4}\right)\frac{P_x^{2}}{\pi^{3}R_{z}^{2}R_{\eta}R_{x}^{2}R_{y}^{2}}\nonumber\\
+\left(\frac{1}{2}G-P_{x}\frac{d x_{0}}{d\tau}\right)\frac{1}{\sqrt{2\pi^{3}}R_{z}R_{\eta}R_{x}R_{y}}.\label{eom}
\end{gather}
Note that the strength of the density-dependent gauge field is characterized by the dimensionless
constant $P_x$.

 In the absence of density-dependent gauge field, these equations
describe the dipole and breathing modes of BEC \cite{perez-garcia_Low_1996}, respectively.
The dipole and breathing oscillations in $x$ direction are always coupled, and for quasi-1D case, these are the only two relevant modes. For quasi-2D case,
the gauge field also introduces a nontrivial coupling between the dipole mode in $x$ direction and the breathing mode in $y$ direction, reminiscent of the Hall effect induced by the gauge field, which is crucial for a comprehensive
description of the condensate dynamics. This coupling shows the key distinction between quasi-2D and quasi-1D cases, and plays an important role in the emergence of chaotic dynamics in quasi-2D case.

\section{chaotic dynamics}\label{dynamics}

\begin{figure}
\centering
\includegraphics[width=0.98\linewidth]{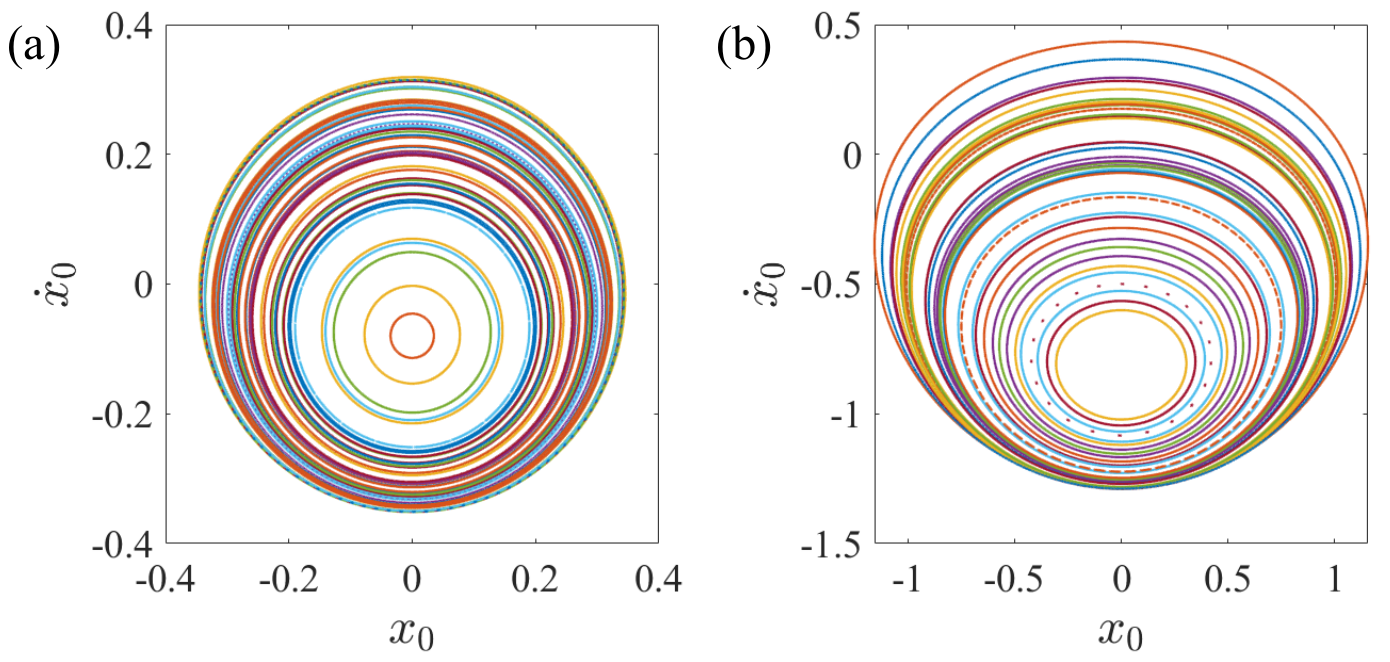}
\caption{2D Poincar\'e section of quasi-1D center of mass motion. Different color denotes
different initial conditions. (a) $P_{x}=0.5$, $E=1.02*E_\mathrm{min}$.
(b) $P_{x}=3$, $E=1.2*E_\mathrm{min}$. The other parameters are $R_y=0.2$, $R_z=0.2$, $\lambda_{x}=1$, and
$G=5$. \label{fig:1D}}
\end{figure}

\begin{figure*}[!htb]
\center{\includegraphics[width=16cm]{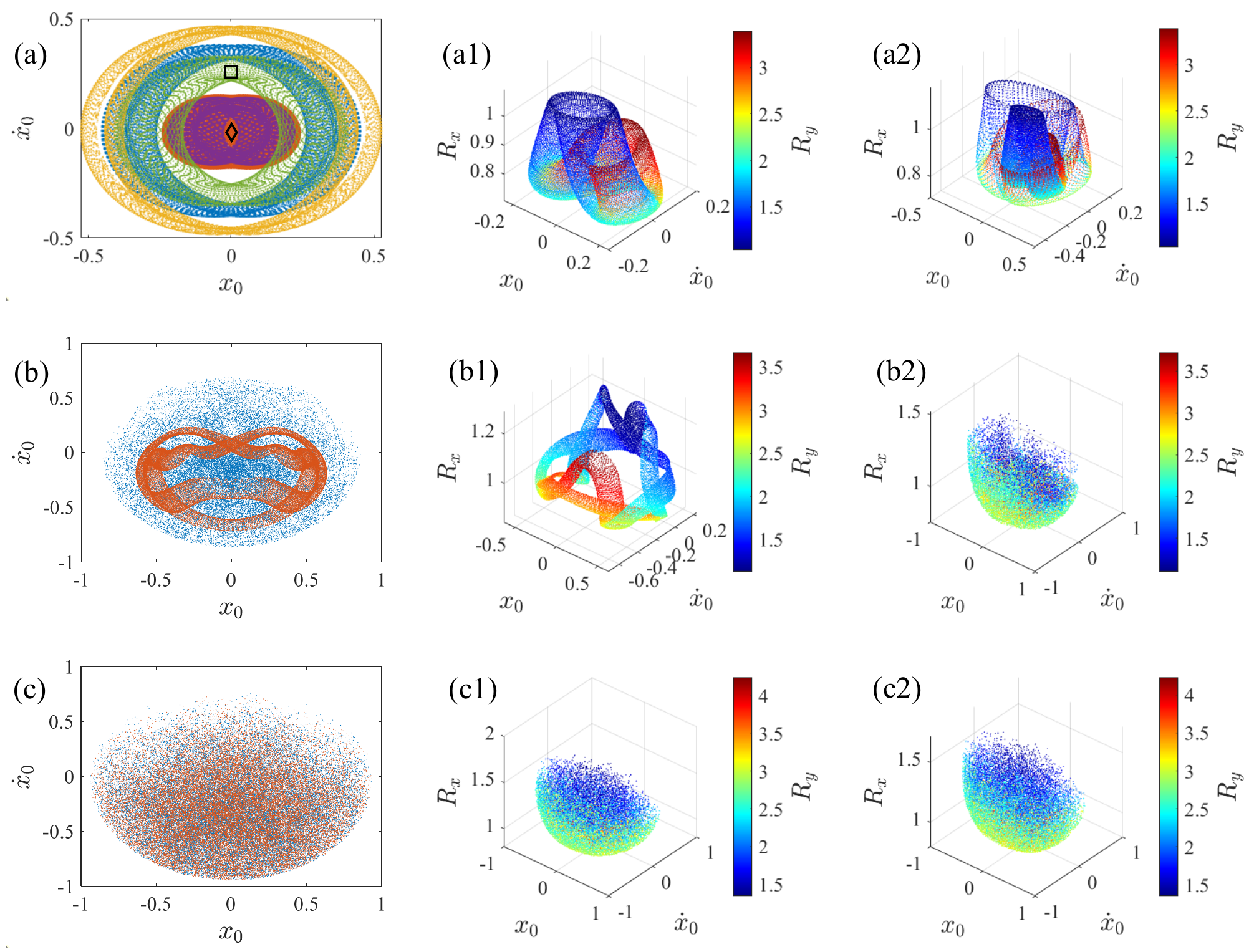}}
\caption{Projected 2D Poincar\'e section on the $x_{0}-\dot{x}_{0}$ plane
and 4D Poincar\'e section for certain orbits. (a) For weak density-dependent gauge field,
only quasiperiodic orbits are present in Poincar\'e section. Different
colors represent orbits with different initial conditions. The original
4D orbits for the central dark red curve (with diamond marker) and outer green curve (with square marker) are shown
in (a1) and (a2), respectively. $P_{x}=0.2$. (b) For moderate
strength of gauge field, the phase space consists
of quasiperiodic (dark red) and chaotic (dark blue) orbits, with their
4D form shown in (b1) and (b2), respectively. $P_{x}=4$. (c) For strong
gauge field, the Poincar\'e
section consists of a chaotic sea, with scattered dark red and dark blue dots representing
two chaotic orbits in (c1) and (c2) with different initial conditions.
$P_{x}=10$. The other parameters are $R_z=0.2$, $\lambda_{x}=1$, $\lambda_{y}=0.5$,
$G=5$, and $E=1.3*E_\mathrm{min}$. \label{fig:2D-compare}}
\end{figure*}

Before solving the coupled nonlinear equations, we first note that the
original Hamiltonian does not change with time, so the total energy is conserved,
\begin{align}
\frac{E}{N} & =\frac{1}{4}\sum_{\eta=x,y}\left[\frac{1}{R_{\eta}^{2}}+\dot{R}_{\eta}^{2}+2\dot{\eta}_{0}^{2}+\lambda_{\eta}^{2}\left(R_{\eta}^{2}+2\eta_{0}^{2}\right)\right]\nonumber\\
 & +\frac{G}{4\sqrt{2\pi^{3}}R_{z}R_{x}R_{y}}+\frac{1}{4}\left(\frac{2}{3\sqrt{3}}-\frac{1}{4}\right)\frac{P_{x}^{2}}{\pi^{3}R_{z}^{2}R_{x}^{2}R_{y}^{2}},
\end{align}
where an unimportant constant $W+\hbar\Omega/2$ is dropped out.
Due to the conservation of total energy, when computing the Poincar\'e
sections in the following, the total energy serves as a constrain for reducing the
dimension of phase space. In addition,
we first calculate the lowest energy $E_\mathrm{min}$ at given parameters, and then use it
as the energy unit.

For small amplitude oscillation, i.e., in the linear response regime,
these equations can be linearized to give the frequency of
coupled motion.  
To study oscillations of larger amplitude, or higher
energy modes, one has to numerically solve these coupled equations
to obtain the center of mass motion, along with the width motion of the condensate.
 The density-dependent gauge field couples these two kinds of
motion, making the overall dynamics complicated. Here the spatial dimension of the underlying dynamics is crucial. 

For comparison, we first consider the quasi-1D case. The condensate is tightly confined
both in $y$ and $z$ directions, with their widths fixed to be the same $R_y=R_z$.
The dynamical variables
are then $x_{0}$, $\dot{x}_{0}$, $R_{x}$ and $\dot{R}_{x}$. As
the dimension of the phase space is four, by fixing the total energy
as well as another dynamical variable, e.g., $\dot{R}_{x}=0$, we obtain a 
conventional 2D Poincar\'e section
in the $x_{0}-\dot{x}_{0}$ plane. 
The Poincar\'e section for two typical strengths of density-dependent gauge field is shown
in Fig. \ref{fig:1D}. One sees that whether the density-dependent
gauge field is weak (Fig. \ref{fig:1D}a) or strong (Fig. \ref{fig:1D}b),
there are only regular dots and closed curves in the Poincar\'e section, corresponding
to periodic and quasiperiodic orbits. These quasiperiodic motion may result from
the irrational frequency ratio of dipole mode over breathing mode.

For the quasi-2D case, since the density-dependent gauge field has been fixed in $x$ direction,
 the center of mass motion in $y$ direction becomes decoupled, as evident from Eq. \ref{eom}, and the remaining dynamical variables are $x_{0}$, $\dot{x}_{0}$,
$R_{x}$, $\dot{R}_{x}$, $R_{y}$ and $\dot{R}_{y}$. The phase space
is six-dimensional, and thus by fixing the total energy along with $\dot{R}_{x}=0$,
the Poincar\'e section is four-dimensional. 
In principle, one can also add two other constrains, e.g., 
by additionally fixing both $R_x$ and $R_y$,
to obtain a 2D Poincar\'e section, which however proves to be numerically quite
challenging because of sparsity of those intersection points.
In the present study, we use a 2D projection to illustrate the overall feature of
the 4D Poincar\'e section, in conjunction with 4D visualization of certain
orbits, similar to methods previously introduced elsewhere \cite{lukes-gerakopoulos_Dynamics_2016}. 
The 2D
projection is on the $x_{0}-\dot{x}_{0}$ plane, and the 4D
visualization of certain orbits is achieved via a three-dimensional
plot of $x_{0}$, $\dot{x}_{0}$, and
$R_{x}$, with $R_{y}$ represented by color. 

Figure \ref{fig:2D-compare} shows the Poincar\'e section using the
method described above, including its original 4D form and 2D projection on the $x_{0}-\dot{x}_{0}$ plane. Only limited number of initial conditions are chosen in
computing the Poincar\'e section,
as those curves in projected 2D Poincar\'e section can overlap
with each other, messing the Poincar\'e section for large
number of initial conditions.
This feature is in stark contrast with conventional 2D Poincar\'e section, where regular and chaotic orbits do not
overlap with each other.
When the density-dependent
gauge field is weak, the projected 2D Poincar\'e section (Fig. \ref{fig:2D-compare}a)
consists of various distorted closed curves, which are originally distorted
tori in the 4D Poincar\'e section (Fig. \ref{fig:2D-compare}(a1),(a2)).
So we conclude that these closed curves in projected 2D Poincar\'e section correspond to
quasiperiodic orbits. 
Figure \ref{fig:2D-compare}(a1) and (a2) show the original 4D Poincar\'e
section for two different initial conditions, corresponding to the
central dark red (with diamond marker) and outer green curves (with square marker) in Fig. \ref{fig:2D-compare}a, respectively.
With the increase of the strength of gauge field,
chaotic orbits eventually appears, as shown by the scattered points in
the chaotic sea of Fig. \ref{fig:2D-compare}(b2), and its 2D
projection in Fig. \ref{fig:2D-compare}b (dark blue dots). Note that once the center of mass motion
becomes chaotic, the motion of widths $R_x$ and $R_y$ are also chaotic. So one can also
choose to project the 4D Poincar\'e on other 2D planes, and similar 2D chaotic sea will
be observed.
In this parameter regime, the Poincar\'e section consists of coexisting quasiperidic
(see Fig. \ref{fig:2D-compare}(b1), corresponding to dark red dots in Fig. \ref{fig:2D-compare}(b)) and chaotic orbits.

For sufficiently strong density-dependent gauge field, e.g., in the parameter
regime shown in Fig. \ref{fig:2D-compare}c, we only find chaotic
orbits starting from various initial conditions. Figure \ref{fig:2D-compare}(c1)
and (c2) show the chaotic orbits with two different initial conditions
in the 4D Poincar\'e section, with Fig. \ref{fig:2D-compare}c being
their 2D projections. The whole phase space in the Poincar\'e section
is occupied by chaotic orbits. We also vary the strength of gauge
field, total energy and trap anisotropy, and results show that stronger
gauge field, higher energy and stronger trap anisotropy
all favor the appearance of chaos in the collective motion of BEC. 

\begin{figure}
\centering
\includegraphics[width=0.98\linewidth]{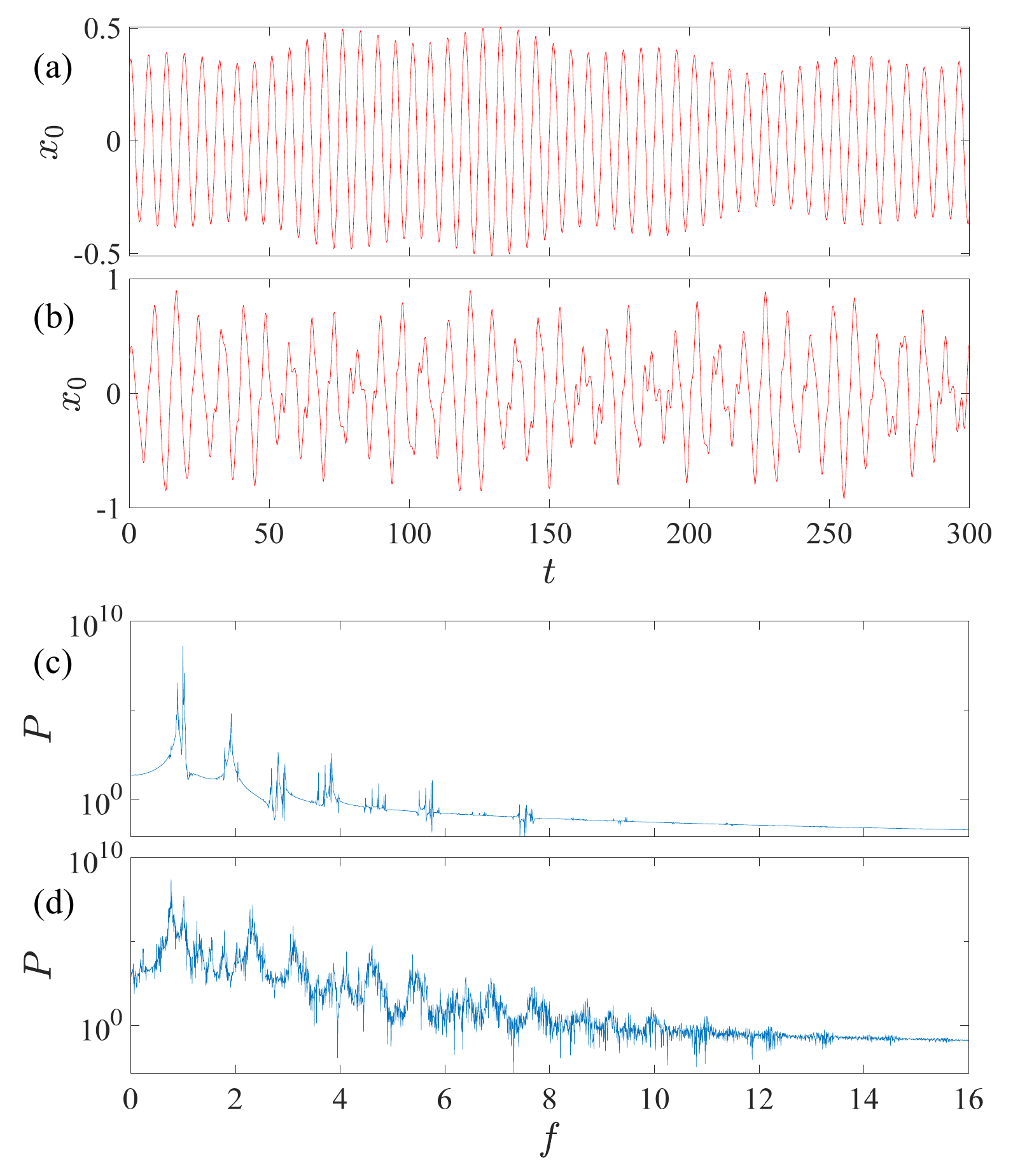} 
\caption{Quasiperiodic and chaotic dynamics of center of mass motion shown in time domain (a and b) and their power spectrums (c and d). (a) and (c) show the quasiperiodic motion, with $P_{x}=0.2$. (b) and (d) show the chaotic motion, with $P_{x}=10$. The frequency $f$ is in the unit of $1/{\nu}$. The other parameters are $R_z=0.2$, $\lambda_{x}=1$, $\lambda_{y}=0.5$, $G=5$, and $E=1.3*E_\mathrm{min}$.
\label{fig:power}}
\end{figure}

To further confirm the nature of these chaotic orbits, we
choose a typical initial condition in the chaotic sea of Fig. \ref{fig:2D-compare}c, 
and show the
time evolution as well as the power spectrum of one typical chaotic
orbit in Fig. \ref{fig:power}(b,d). The chaotic dynamics is manifested by the broad and structureless feature in 
power spectrum. For comparison, the time evolution and power spectrum of a typical
quasiperiodic orbit is also shown in Fig. \ref{fig:power}(a,c), with same parameters as Fig. \ref{fig:2D-compare}a.
The distinction between these two kinds of orbits is clear in both time evolution and
power spectrum, consistent with our judgment based on Poincar\'e section. 

The chaotic dynamics can be qualitatively understood as the interplay between three oscillations, namely, the dipole and breathing oscillations along the direction
of gauge field, and the breathing oscillation in perpendicular direction.
The former two modes are coupled through the
current nonlinearity \cite{edmonds_Simulating_2013}, while the dipole and breathing modes
in perpendicular directions are coupled by the 
interaction induced Hall effect, only present for system of dimension two or higher. This explains the dramatic difference between quasi-1D and quasi-2D dynamics.

The major challenge towards the experimental observation of the chaotic dynamics of BEC 
predicted here is the relatively small difference between different scattering
lengths, and therefore the density-dependent gauge field is relatively
weak. Nevertheless, in principle one can employ the Feshbach resonance technique \cite{chin_Feshbach_2010}
to enhance the difference between different scattering lengths. The chaotic
dynamics of BEC will be observed by measuring either the center of mass or the width
motion of BEC in a harmonic trap. 

\section{conclusion}\label{conclusion}

In summary, we have studied the effect of density-dependent gauge
field on the collective dynamics of a harmonically trapped BEC. The center of mass dynamics
depends sensitively on the spatial dimension of the condensate. For
quasi-1D BEC, we only find quasiperiodic motion, whether the 
gauge field is weak or strong. In contrast, for quasi-2D BEC and strong 
gauge field, we find that the center of mass motion can become chaotic.
This is also true for the dynamics of condensate width. The chaotic behavior
are induced by the interplay between the dipole oscillation in the direction of gauge field
and the breathing oscillations both along and perpendicular to that direction.
The coupling between dipole and breathing oscillations in perpendicular direction
can be understood as a Hall effect induced by the gauge field.
 We confirm the chaotic dynamics by computing the 4D Poincar\'e section, its projection
on 2D plane, and the power spectrum. We find strong density-dependent
gauge field, trap anisotropy and high energy are favorable
for the observation of chaotic dynamics. Our findings deepen our understanding
of the effect of dynamical gauge field in cold atomic gases, especially
on the nonlinear dynamics of BEC.

\section{acknowledgments}

We thank Qiong-Tao Xie and Biao Wu for helpful discussions. L.C. is
supported by the Science Foundation of Guizhou Science and Technology Department (Grants No. QKHJZ[2021]033 and No. QKHJZ[2018]1178), and the Science Foundation of Guizhou Provincial Education Department (Grant No. QJHKYZ[2017]087). Q.Z. is supported by the National Natural Science Foundation
of China (Grant No. 12004118), the Guangdong Basic and Applied Basic Research 
Foundation (Grants No. 2020A1515110228 and No. 2021A1515010212), and the Science and Technology Program of Guangzhou (Grant No. 2019050001).

\end{document}